# A New Method of 3D Magnetic Field Reconstruction

R. B. Torbert[1,2], I. Dors[1], M. R. Argall[1], K. J. Genestreti[2], J. L. Burch[2], C. J. Farrugia[1], T. G. Forbes[1], B. L. Giles[3], R. J. Strangeway[4]

1. University of New Hampshire, Durham, NH, USA
2. Southwest Research Institute, San Antonio, TX, USA
3. NASA Goddard Space Flight Center, Greenbelt, MD, USA
4. University of California, Los Angeles, CA, USA

Corresponding Author
R.B. Torbert, 406 Morse Hall, University of New Hampshire, Durham,NH 03824 (roy.torbert@unh.edu)

Abstract

A method is described to model the magnetic field in the vicinity of constellations of multiple satellites using field and plasma current measurements. This quadratic model has the properties that the divergence is zero everywhere and matches the measured values of the magnetic field and its curl (current) at each spacecraft, and thus extends the linear curlometer method to second order. It is able to predict the topology of the field lines near magnetic structures, such as near reconnecting regions or flux ropes, and allows a tracking of the motion of these structures relative to the spacecraft constellation. Comparisons to PIC simulations estimate the model accuracy. Reconstruction of two electron diffusion regions show the expected field line structure. The model can be applied to other small-scale phenomena (bow shock, waves of commensurate wavelength), and can be modified to reconstruct also the electric field, allowing tracing of particle trajectories.

## 1. Introduction

Measurements and models of the magnetic field are commonly studied in the extensive space physics literature. The magnetic field **B** is the predominant reservoir of energy available for acceleration of particles. Particle trajectories and energization processes are greatly influenced by the magnetic field and its topology. For these reasons, magnetometers are the one of the most common instruments in ground and space missions (see Kivelson and Russell, 1995). Over the past several decades, multiple-satellite mission designs (e.g., the International Sun-Earth Explorers, Cluster and Magnetospheric Multiscale (MMS)) have been employed as they allow for approximate determination of the topologies of magnetic boundaries, which facilitates more complete analyses of related plasma phenomena.

The latest of these missions, MMS, launched in 2015 (Burch et al., 2015) has targeted magnetic reconnection in boundary regions of the magnetosphere. The magnetic topology plays a key role in energy conversion, fast flow acceleration and energetic particle production, which are characteristic of reconnection. Reconstructions of **B** and its streamlines not only provide a more complete picture of fundamental features of reconnection regions, but also allow simple recognition of where the spacecraft are located and how they are moving relative to magnetic structures.

There have been many approaches to reconstructing **B**: Grad-Shafranov techniques introduced by Sonnerup and Guo (1996), with the addition of flow parameters by Sonnerup and Teh (2008); using constraints imposed by magnetohydrodynamics (MHD) (Sonnerup and Teh, 2009); and reconstructions under some simplifying topologies and the constraints of electron MHD (Sonnerup et al., 2016). Using a linear approximation of $\nabla\mathbf{B}$ (Dunlop et al., 1988), Shi et al. (2005) developed a method for estimating the motion of magnetic structures relative to spacecraft; and Denton et al. (2016) have refined this technique to investigate motions of the MMS tetrahedron configuration. However, a weakness of this method is that the experimental $\nabla\mathbf{B}$ commonly does not have zero trace (i.e., $\nabla\bullet\mathbf{B}\neq0$) and the components corresponding to the curl are those of an average assumed for the barycenter of the spacecraft: i.e the current density within the spacecraft tetrahedron is assumed to be uniform.

One of remarkable advances of the MMS mission is the very high fidelity of the current measurements using only particle data (Pollock, et al., 2014; Phan et al., 2016). The configuration of eight spectrometers per spacecraft that simultaneously measure flows in opposing directions for electrons every 30ms and ions (150ms) is a significant asset for this success. Furthermore, the magnetometers (Russell, et al., 2014), assisted with independent measurements of the magnitude by the Electron Drift Instrument (EDI) (Torbert et al., 2014), have provided one of the most accurate measurements of **B** ever acquired by high-altitude spacecraft, with an accuracy of ≤0.1 nT. Using a "modified" curlometer, which employs both time and spatial variations of **B** to estimate current, Torbert et al. (2017, see Figure 1) showed that the particle data matched the magnetic variations at the highest cadence available on MMS within an electron diffusion region (EDR), where the current is far from uniform.

In this letter, we propose a new method where we use these accurate measurements at each spacecraft to produce a local, basically quadratic model of **B** that exactly matches the measurements of **B** and ∇x**B** and has zero divergence everywhere. This technique extends the linear curlometer method to second order and will allow better estimates of both local field line topology and the motion of magnetic structures by the spacecraft.

## 2. Model Description

One of Helmholtz's theorems states that a three-dimensional, continuously differential vector field is uniquely specified within an enclosed volume by 1) its curl; 2) its divergence; and 3) the normal component over the boundary (Arfken,1985). The divergence of **B** is zero everywhere. Given that the displacement current is negligible, for the time scales modeled here, the curl is the current (times $\mu_0$) measured by the particle instrumentation on the four MMS satellites, but of course only at four vertices of the constellation tetrahedron, where the values of **B** are also determined. Thus, given that the curl and normal component are not fully specified, the complete vector field obviously cannot be modeled. However, if the spatial variation is restricted, a model can be reconstructed that is the simplest (most slowly varying) possible one that is consistent with the data. There are several approaches to this problem (interpolations along the boundary and within the volume, using Helmholtz constructions of the field), but the most straightforward is to Taylor expand the field at some convenient point within the MMS tetrahedron (here taken as the barycenter), and truncate the series when there is a sufficient number of coefficients. The result will be nearly a quadratic expansion (by "nearly," see below) for each component (j):

$$B_j = B_{0j} + \sum_k x_k (\partial_k B_j)_0 + ½\sum_{k,l} x_k x_l (\partial_k \partial_l B_j)_0 \qquad (1)$$

where each of the model coefficients [ $B_{0j}$, $(\partial_k B_j)_0$, $(\partial_k \partial_l B_j)_0$ ] are referenced to the expansion origin(0), and the $x_k$ are the components of the position of the field point referenced to that location. Without the quadratic term, such an expansion will replicate the normal curlometer method (Dunlop,1988) when its 12 free parameters (three $B_{0j}$ and nine $(\partial_k B_j)_0$ are determined using the 3*4 measured components of **B**. Extending this expansion to the quadratic term allows us to model the magnetic field with our knowledge of the current at each spacecraft.

Given that the quadratic coefficient is symmetric in the k-l indices, there are 3(components) * (1 + 3 + 6 coefficients), or 30 unknowns in this expansion. However, the divergence-free requirement on **B** implies that the trace of the linear term is zero and the gradient of the trace is also zero (constraining the quadratic coefficients), and therefore four of these unknowns are determined, reducing the number to 26. We have four spacecraft observations of both **B** and **J,** providing 24 elements of data. We thus need additional constraints on the expansion. We obtain these by using a minimum variance analysis (Sonnerup,1998) to produce a local LMN coordinate system (where M is the minimum variance direction and N is the maximum variance direction) and demand that the three $\partial_M \partial_M B_j$ terms be zero, since there is little variation in this direction. Given that there are now 23 parameters for 24 measurements, the problem would seem to be over-constrained. However, since **J** is computed from the curl of **B** , the model automatically delivers a divergence-free current, whereas the data will not be so: not only because (1) there are errors in the current measurement itself, but also because (2) the current is measured at four separated points, and there is no requirement that the *linear approximation* of the gradient tensor of **J**, using these separated points, be traceless. Thus, if the data could be constrained so that the trace of $\nabla$**J** is zero, there would be 23 data values for 23 parameters, resulting in a unique solution.

It is tempting to resolve this issue, the non-zero divergence of **J** data, by devising some method of adjusting that data itself, but analysis within regions of strong current usually shows a strong spatial and non-linear variation such that, in fact, where the authors want to model the changing **B** field (in reconnection diffusion regions), reason (2) above dominates: the errors are less than the variation over the tetrahedron, as is clearly seen in Figure 1. Another possibility is to use fewer expansion parameters and do a least-squares fit to the data to determine a less varying model, which, however, does not match the measurements at each satellite (R.Denton, private communication). This approach may produce a better model extrapolated farther from the spacecraft, because some of the quadratic terms (especially cross-terms in the M-derivatives) will give spurious results at large distances. However, this first letter describes the procedure to produce a higher fidelity fit within, or very close to, the tetrahedron itself, for the future purposes of modeling particle trajectories within and near the tetrahedron, as described below.

In this case, to produce a model where the variation in **J** is not completely linear, an additional cubic term is required to produce a fit, given the 24 independent measurements of **B** and **J**. If we impose requirements that: 1) no terms may have more than a linear dependence in the M-direction, consistent with the approach above; 2) only a single term be added, to be varying as simply as possible; and 3) the divergence of the field be constant (in the case of **B**, namely zero, but see below ), then, given the symmetry in partial derivatives, careful examination of all the combinations shows that there are only eight possibilities to add a single additional cubic term, $(\partial_i\partial_k\partial_l B_j)_0$ ,to equation (1) above: namely, [iklj] = [1132], [ 1123], [ 3321], [1332], [1113], [3331], [3332], or [1112], where 123 = NML in the above coordinate system. In principle, any one of these will give an exact fit to the data. However, since our objective is to find the smoothest (least varying) model fit over the tetrahedron, the coefficients with the largest scaling length (favoring the smallest cubic coefficient) are preferred. In practice, a solution is obtained for each of the eight cubic terms, and a weighted average of all of the solutions with scale lengths within a factor of four of the maximum is computed as the final model. The final result usually involves 2 to 3 of the possible solutions and allows a continuous time evolution of the field. In principle, a solution can be obtained for every time step where there is a reliable current measurement (on MMS, every 30ms, usually). Since the result is a linear combination of exact solutions, the final model also has zero divergence for **B** everywhere; it matches the observed **B** and **J** at each spacecraft, and varies spatially as slowly as possible, and very nearly quadratically throughout the tetrahedron. We call this solution the "25 parameter" fit: 23 from Equation (1), one cubic term, plus a 25$^{th}$ parameter which is a constant divergence. For **B**, of course, this parameter is identically zero. However, as described below, there are mathematical reasons to retain this quantity as a constant parameter. The matrix that results, and which must be inverted to obtain the 25 parameters, is given in (**S1**).

Since the model is basically only quadratic, it is critical, before solving for coefficients, to average the data to the appropriate time scale, corresponding to an appropriate spatial scale with an assumption about the average speed of structures past the spacecraft. Clearly waves, for example, with wavelengths much smaller than the spacecraft separation, cannot be replicated in a model that uses data at four separated points. In the examples below, timing methods were used to estimate structure velocity, and then the data were averaged on a time scale corresponding to a spatial scale of about half the tetrahedron size, providing an approximate three-point fit to a quadratic.

## 3. Comparisons with Model Simulations

As an initial test of the procedure, in the absence of a complete map of the field throughout a real tetrahedron, we have compared our model results with those obtained in PIC simulations of reconnection. Data on the magnetic field and currents from the simulation of Nakamura (2018) were obtained along four tracks, with separation corresponding to those of MMS on 11 July 2017 as reported by Torbert (2018). The data were averaged over time as described above. This simulation was 2.5D, thus the PIC data are constant in the M direction. However, the reconstruction algorithm was not informed or adjusted for this. Nevertheless, the coefficients that were determined all showed negligible values for M-derivatives. Plots of the comparison between simulation data and reconstruction in Figure 2 show excellent agreement in both the magnitude of the field (2b) and the direction (2d) in a volume about twice that of the tetrahedron. Figure 2c, showing the $B_N$ component, clearly shows the effect of quadratic terms when farther out from the center of the spacecraft constellation.

4. Two Example Electron Diffusion Regions (EDRs).

Two of many cases that have been reconstructed with this method are the dayside asymmetric reconnection event of 16 October 2015 (Burch, et al., 2016) and the magnetotail symmetric reconnection event of 11 July 2017 used above (Torbert, at al., 2018). Figure 3a shows the results of the 3D reconstruction when MMS was very near an EDR on 16 Oct 2015 at the magnetopause. The field vectors are computed every 30ms on a cubic grid with 2 km spacing, whereas the spacecraft (colored diamonds) are separated by ~10-15 km. Only a few representative field lines, computed from this field array, are shown, four of which go through the location of spacecraft. Although they are three dimensional and have a M-component, the field lines are projected onto a LN plane, using the LMN system given in (Burch, 2016). Using the magnetic, electric, and current field data (in 3b) and the electron distributions (in 3c), Burch confirmed that MMS4 was south of the EDR, and that MMS2 and 3 were north of the EDR at this time, which corresponds to the peak in the **J•E** dissipation. This location is clearly confirmed by the reconstruction. The advantage of the quadratic reconstruction is apparent in 3a: even though all four measurements of **B** at this time are in the same direction (see top panel of 3b), the current (∇x**B**) demands that the field reverse just beyond the tetrahedron (at larger N position) and that an X-line lies near MMS4. The reconstruction of this and other EDRs gives a definitive visual confirmation of the topological changes implied by reconnection.

The second example (Figure 4) uses data from the 11 July 2017 encounter with an EDR in the magnetotail (Torbert, 2018). The authors there showed that the MMS constellation traversed an EDR in the earthward and (meandering) northward directions while remaining near the neutral plane (where $B_L \sim 0$). Figure 4a shows six seconds of magnetic, electric, and current field data consistent with that interpretation. In Figure 4b, the constellation is seen earthward and southward of the X-line, not yet fully within the earthward electron outflow jet. Later, in Figure 4c, the spacecraft have moved closer to the neutral plane and are now fully within the earthward jet. The reconstructions help visualize the position of MMS relative to the EDR structure, and also new features that may be missed in the data: the field lines that terminate in the N-L projection at 3.36s do so because they exit the cubic reconstruction matrix in the M direction before coming to the N-L boundary. This is the effect of the now much larger $\mathbf{J}_L$ current that twists the field in the outflow jet and creates the "Hall" magnetic field (Sonnerup, et al, 1979). The reconstruction software developed for this method allows 3D visualizations of these field lines where such effects are more readily seen. In a movie of the entire six seconds of Figure 4, the motion, sometimes back-and-forth, of the field line structure can be seen. In this movie, sometimes quite convoluted topologies are seen, but it should be emphasized that each solution is the *simplest* one consistent with the data: they could be, and probably are, often even more complicated!

4. Conclusions and Future Work.

The reconstructions described here are done over time intervals of typically 0.1 s, consistent with the averaging procedure above, but can in principle be done on at the fastest cadence for the current: on MMS, that is 7.5ms. The averaging, however, reduces the statistical fluctuations in the current measurement. Besides the comparison with simulations above, we have assessed the sensitivity to errors in the current to the generated topology (see S3): errors of 5-10 % in the current have only a small effect owing to the fact that the field itself is fixed at each spacecraft and the errors in **B** are very small for MMS (~0.1 nT). As is seen before and after the EDR in the supplemental movie (where there is very little current and this percentage error is not unexpected), the field lines are very regular and this percentage error has little effect. Where the currents are large, within EDRs such as seen between 12:59:14.1-14.2s in Figure 1, this error is well within the capabilities of the FPI instrument. Work is continuing to analyze how robust the model is to these errors. Future work by R. Denton (private communication) is also exploring the use of least-squares fitting to models with fewer parameters which may result in models of the topology valid at larger distances from the tetrahedron.

Besides the need to determine the magnetic topology around MMS, there were two other motivations for this model. The first follows from another Helmholtz theorem that states that the **∇B** matrix at any point can be decomposed into three parts: one, the divergence (trace); the next, the curl (anti-symmetric part); and a third traceless symmetric part which encodes the values of the normal component of a surrounding volume (consistent with the previously cited theorem). We know the first part, and measure directly the second part at each spacecraft; the model provides an estimate of the third consistent with the observations of **B** and **J** at the other three spacecraft. Now, with a matrix valid at *each* spacecraft, a modification of the Shi (2005) method may produce a more reliable estimate of the motion of structures and how the configuration is changing with time, consistent with the time series of reconstructions.

The second additional motivation was to model particle trajectories in EDRs. The gyroradius of the electrons of relevant energies are of the order of, or usually larger, than the spatial variations, and, of course, the flow is not frozen-in, so that the field lines by themselves do not indicate plasma motion. However, MMS has successfully flown an accurate 3D electric field (**E**) measurement, again calibrated with EDI (Torbert,2014; Lundqvist, 2014; Ergun, 2014). For this field, we know the values of **E** and its curl ( $-\partial \mathbf{B}/\partial t$) at four points, and the measurements of **E** are sufficiently accurate to estimate the divergence using the linear curlometer technique. This divergence is certainly not constant, but comparison to simulations shows that it is varying over a scale not much smaller than the spacecraft separation in some of the EDRs that MMS has encountered. We can thus compute a 25-parameter fit for **E** in the same manner as **B**, where this $25^{th}$ parameter is now the non-zero, but constant divergence measured. The model then produces a self-consistent solution of quasi-static Maxwell's equations (no displacement current, i.e. the divergence of **J** is zero) for **B** and **E** around the tetrahedron. Although the assumption of constant divergence limits the spatial applicability of the **E** solution, initial work shows the model has promise for understanding the acceleration of electrons around the EDR and also results in a robust and self-consistent calculation of the terms of Poynting's theorem (electromagnetic energy flux and **J•E**) for studies in the vicinity of the tetrahedron.

Use of these many aspects of this quadratic model for fields around the MMS tetrahedron promises to guide the interpretation of the motion of structures past the spacecraft constellation and further our understanding of acceleration in the very complicated dynamics of reconnection. In addition, the model can be used in the same manner for many other space physics phenomena where the scale sizes are appropriate for the expansion, such as bow shock encounters, and with plasma waves whose wavelengths are comparable to the spacecraft separation.

Acknowledgements.

We have benefitted from conversations with R.E. Denton and others on the successful MMS team, whom we thank for such wonderful data (available at https://lasp.colorado.edu/mms/sdc/public/) and NASA support via contract NNG04EB99C.

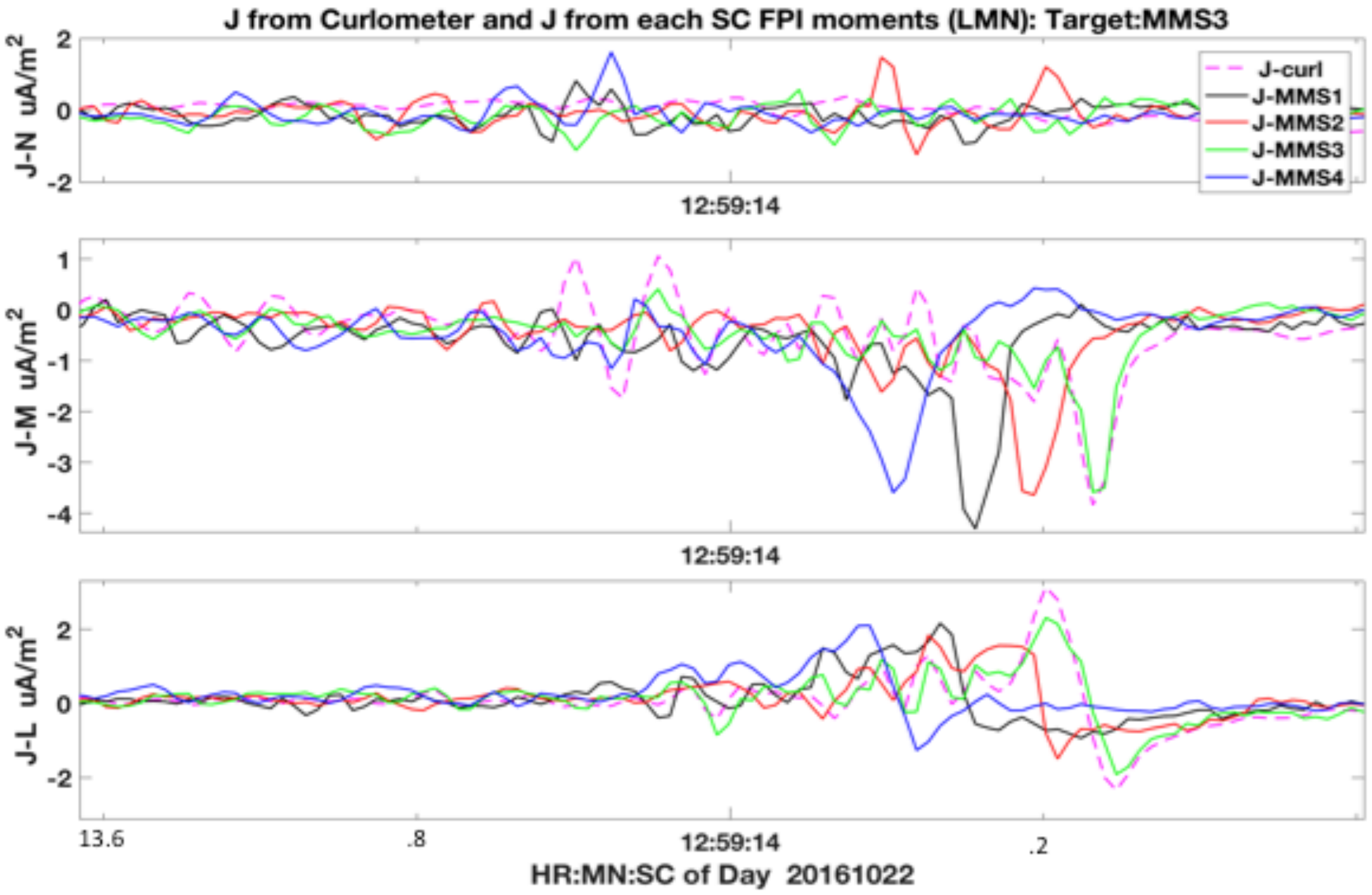

**Figure 1.** Plots of the three components of the current (reproduced from Figure 13, Torbert, et al.,2017), in the LMN coordinate system determined there, as computed from two different techniques: solid lines, the current from plasma spectrometers on each of the four MMS spacecraft; dotted line: current computed for the single MMS3 spacecraft with magnetic field data only, using the modified curlometer described in that paper. The current structures are shown in that paper to be very narrow, of order 2-4 km (whereas the spacecraft separation is ~7 km), resulting in the clear separation of the current signatures. Thus the conventional linear curlometer result, basically an average of the four MMS current peaks at a given time, is much poorer than shown.

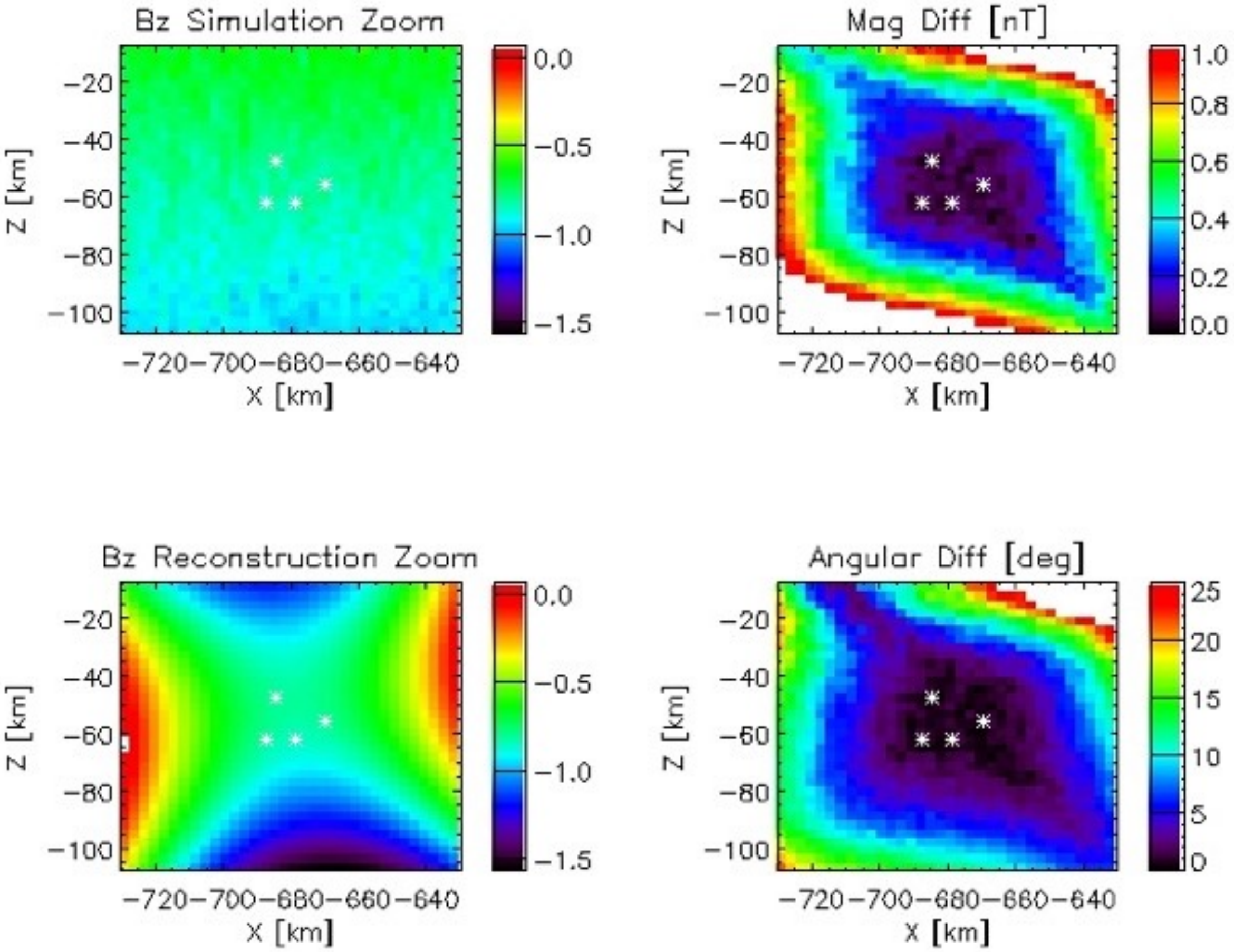

**Figure 2.** Comparison of model results to those near a reconnection diffusion region in a PIC 2.5D simulation by Nakamura (2018). Left panels: the $B_Z$ ($B_N$) component (color coded in nT) of the simulation (upper) and the 3D model (lower) computed from **B** and **J** values at the position of the four MMS satellites (white stars) projected into the X-Z (L-N) plane. Right panels: Differences of the PIC values and the model results of the magnitude (upper) and angular difference (lower, in degrees).

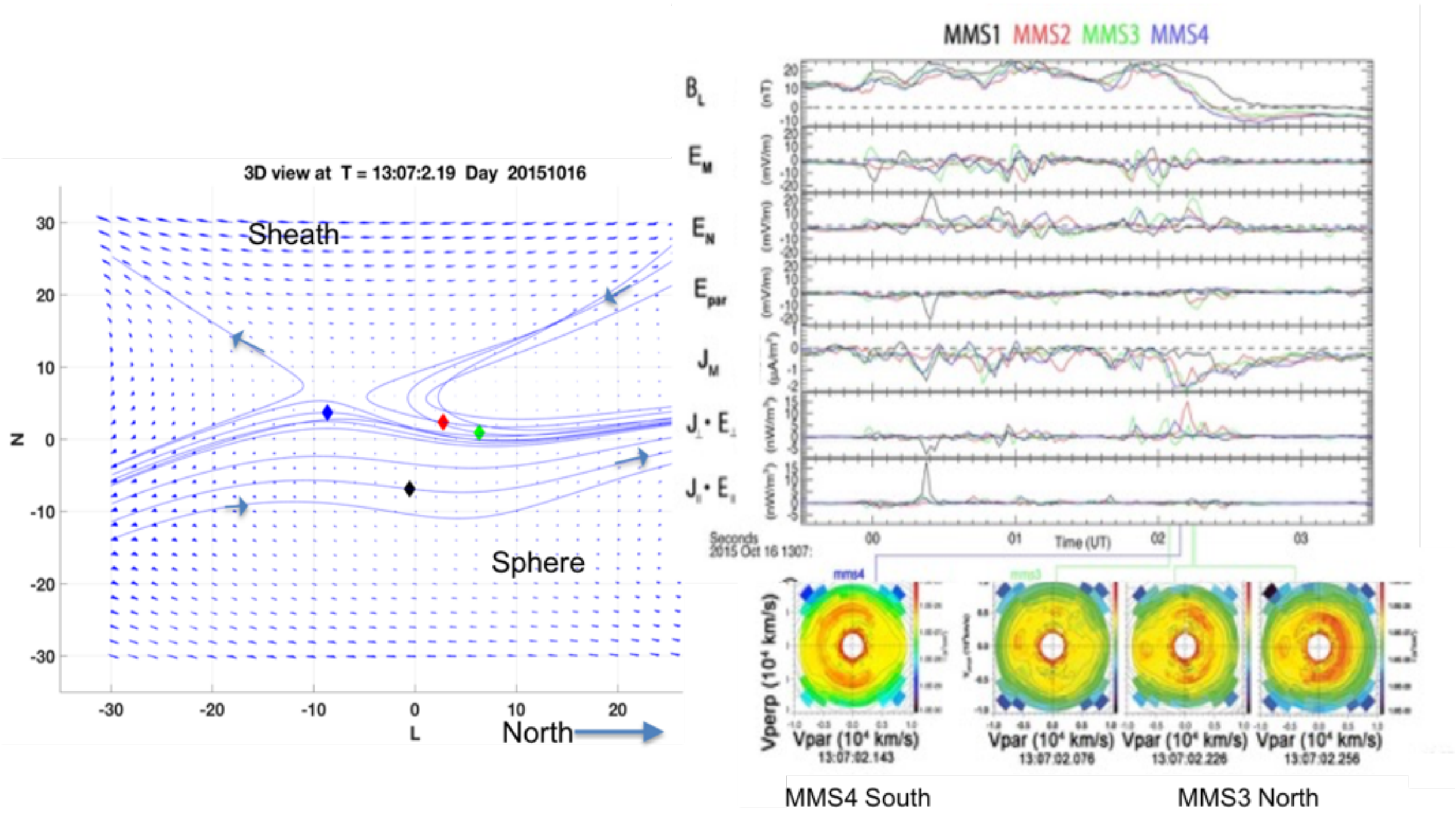

**Figure 3.** Left panel(a): the 3D reconstruction of the magnetic field near an EDR on 16 October 2015. The magnetosheath (magnetosphere) is in the upper (lower) area of the reconstruction. The vector field is computed on a 3D grid with spacing of 2 km, and the view is through this cubic lattice along the M-direction of an LMN coordinate system given in Burch (2016). The field lines are 3D but projected into the L-N plane. The four MMS spacecraft locations are color coded by the pattern given in top right. Upper right panel(b), ~4s of data taken during this encounter on the four MMS spacecraft, from the top: $B_L$, $E_M$, $E_N$, $E_{parallel}$, $J_M$ (the reconnecting current component), $J_\perp E_\perp$ and $J_{||}E_{||}$ indicating that most of the energy is converted from the perpendicular components; the $B_L$ change marks the approach of the EDR at 13:07:02.2s. Bottom right: electron phase space density ($V_{par}$ , parallel to **B**,horizontal axis; $V_{perp}$ , in the $E_{perp}$ direction) from MMS4 and MMS3 at the indicated times, showing electrons streaming left (southward) on MMS4 and right (northward) on MMS3 consistent with their locations in the reconstruction.

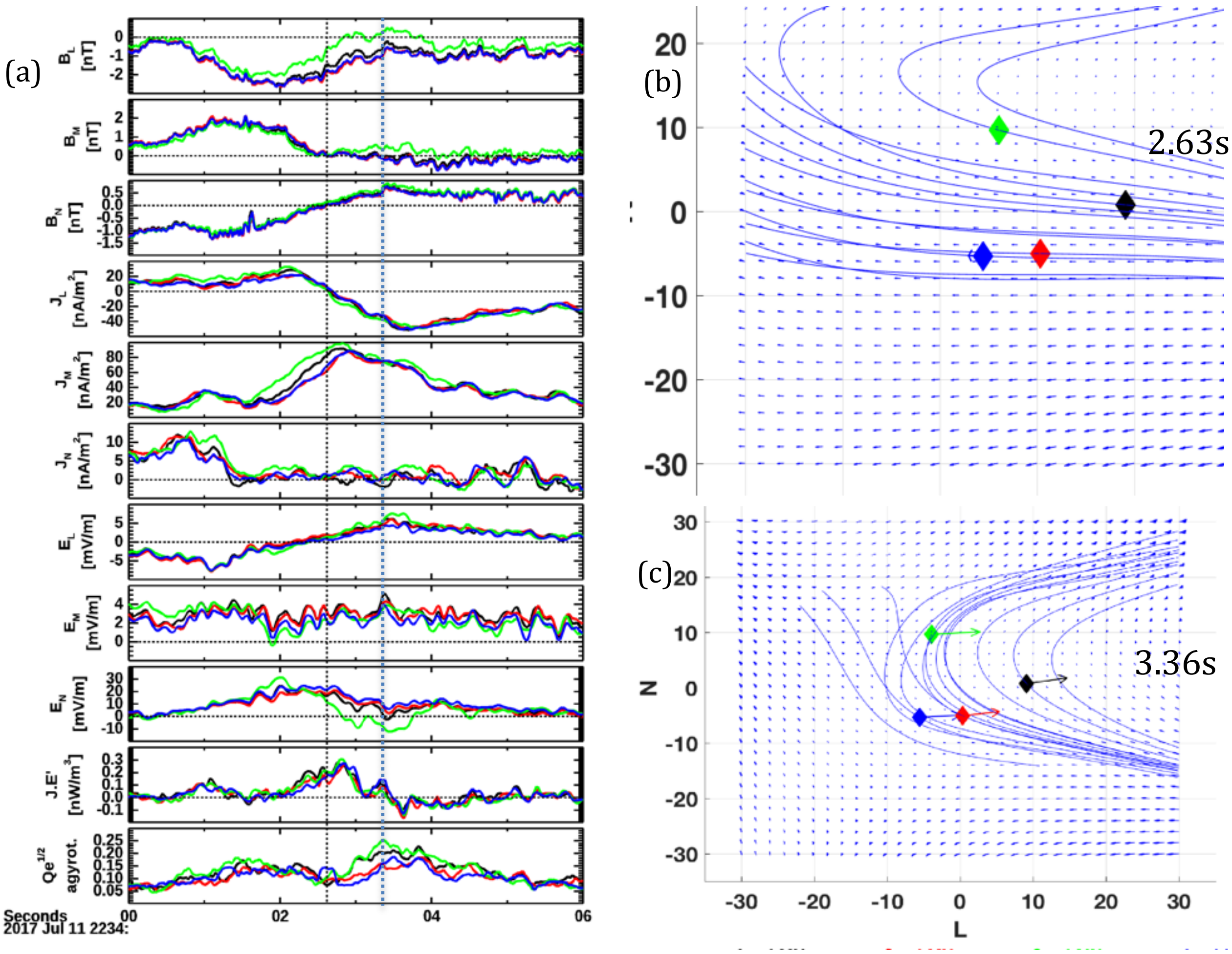

**Figure 4.** Reconstructions made at two different times in the EDR encounter of 11 July 2017 (Torbert,2018). Left panel(a), from the top: $\mathbf{B}_{L,M,N}$ , $\mathbf{J}_{L,M,N}$ , $\mathbf{E}_{L,M,N}$ components during 6s around the times( dotted vertical lines) of the reconstructions; the electromagnetic energy conversion in the electron flow frame(**J•E'**); the square root of the Quisdak agyrotropic parameter, Qe. Upper right panel(b), reconstruction at 22:34:2.63 seconds projected into the LN plane, showing the approach of the four MMS spacecraft toward the neutral plane ($B_L$ = 0) but not yet fully in the electron exhaust of the EDR to the left (<~ - 30 km). Lower right panel(c), MMS at 3.36s, now fully in the earthward electron exhaust at the neutral plane, indicated by the colored arrows which are the L-N projection of the electron bulk flow. There are flow arrows in panel (b), but so small as to be unnoticeable. Some of the field lines terminate as they exit the reconstruction cubic lattice in the M direction before they reach the LN boundary.

Supporting Information for

# A New Method of 3D Magnetic Field Reconstruction

R. B. Torbert[1,2], I. Dors[1], M. R. Argall[1], K. J. Genestreti[2], J. L. Burch[2], C. J. Farrugia[1], T. G. Forbes[1], B. L. Giles[3], R. J. Strangeway[4]

1. University of New Hampshire, Durham, NH, USA
2. Southwest Research Institute, San Antonio, TX, USA
3. NASA Goddard Space Flight Center, Greenbelt, MD, USA
4. University of California, Los Angeles, CA, USA

Corresponding Author
R.B. Torbert, 406 Morse Hall, University of New Hampshire, Durham,NH 03824 (roy.torbert@unh.edu)

## Contents of this file

### Text S1.

The matrix for the computation of the 25 parameters included in the expansion of Equation (1) is described in the Excel file "Torbert_25parameter_matrix.xlsx." The parameters are numbered in row 7, and described in row 5. Both quadratic and the single cubic terms are labelled "Tijk" or "Tijkl", corresponding to the "Terms": $(\partial_i\partial_j B_k)_0$ or $(\partial_i\partial_j\partial_k B_l)_0$ .The parmeters "gbij" label the linear terms $(\partial_i B_j)_0$. This specific matrix gives the values for the $T_{1132}$ cubic version, but all other versions are easily constructed by modification of the 24th parameter given in column AG. The matrix is constructed for a particular configuration of the tetrahedron where the (x,y,z = N,M,L) locations of spacecraft "1" are the values (1X,1Y,1Z), and so forth for all spacecraft. Once the matrix is constructed, the **B** and **J** values in rows 8 through 31, and the divergence term in row 32, can be computed by multiplication with the 25 parameters labelled in row 7. The blank columns marked with an "X" in row 6 are those which are either zero by assumption, ( $\partial_M \partial_M B_j$)0, or determined by the necessity to keep the divergence constant. To first compute the parameters, the matrix is inverted and the data in column B is used to produce the parameters. Thereafter, the **B** and **J** anywhere with location (1X,1Y,1Z) can be computed with the values in rows 8-13. The entries colored in green are a consistency check that the divergence is the constant value of row 32, column B. Likewise, the matrix given in rows 37-45 can be used to compute **∇B** anywhere. Examination of the yellowed colored rows, 37,41, and 45, reveal that the trace of **∇B** is everywhere equal to the divergence specified in row 32.

**Text S2, Figure SF1, and Movie**

The file, "all4mms_plut_reconstruction_mvave_coords_20170711_490.mov, " contains a movie of six seconds during the passage of MMS past the EDR of 11 July 2017, as described in Torbert, et al, (2018). One panel of this movie is shown in Figure SF1. The LMN coordinates are given in GSE as [ 0.9482 -0.2551 -0.1893; 0.1818 0.9245 -0.335; 0.2604 0.2832 0.9230] , slightly updated from that paper. The velocity distributions are given for all four spacecraft, rather than just the two in Figure 4 of the main text. The upper right panel shows the projections of 3D field lines into the LN plane at each time given in the movie.

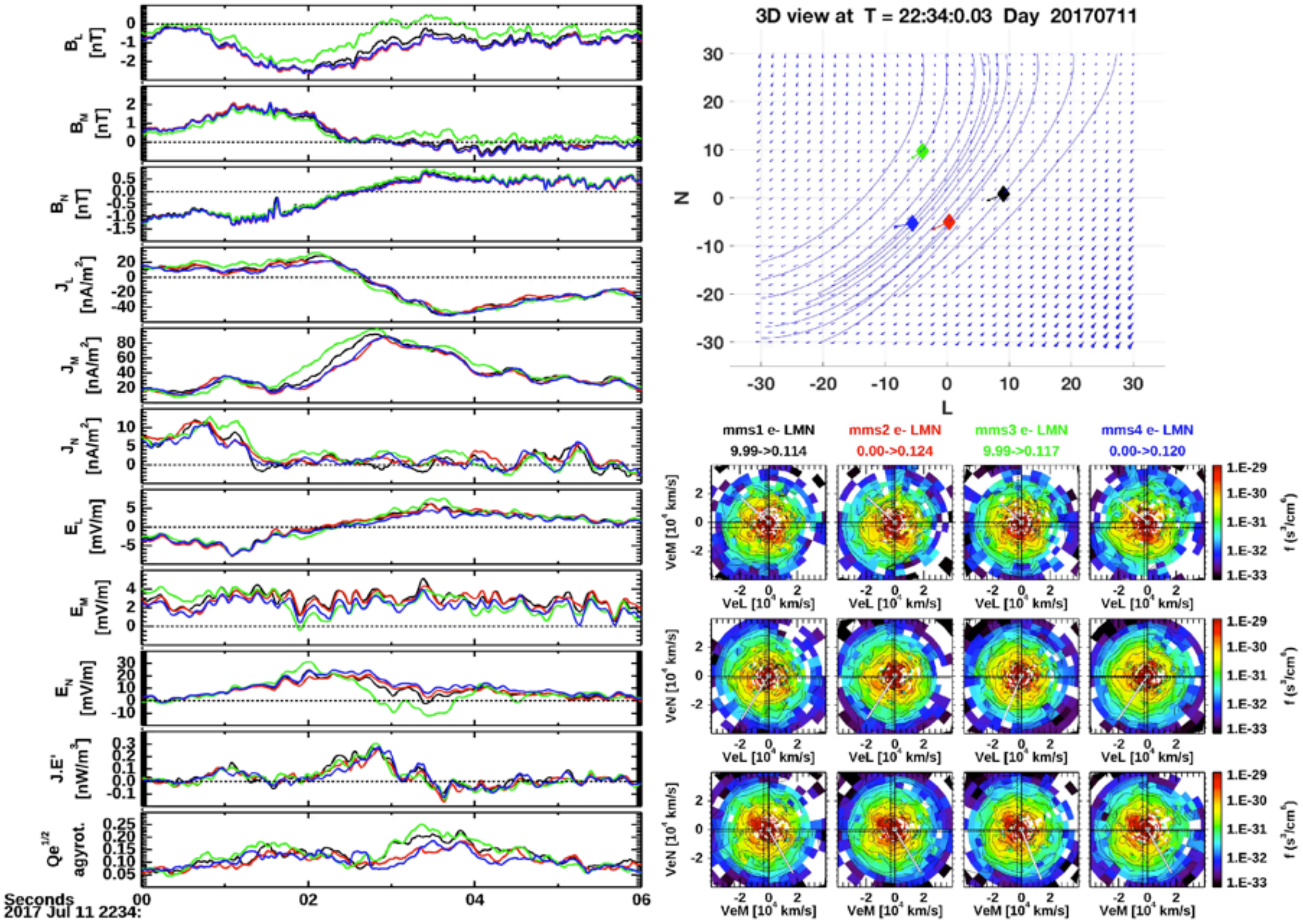

**Figure SF1**

In the left panel, six seconds of $\boldsymbol{B}_{L,M,N}$, $\boldsymbol{J}_{L,M,N}$, $\boldsymbol{E}_{L,M,N}$, $\boldsymbol{J}\cdot(\boldsymbol{E}+\mathbf{v}_e\mathrm{x}\mathbf{B})$, and the Swisdak agyrotropy index on 11 July 2017. Upper right is the reconstruction at the indicated time, with the four MMS spacecraft with the electron bulk flow ( arrows colored according to spacecraft color in lower right panel). Lower right panel: for each spacecraft in a column, the reduced electron distributions along the indicated LMN directions.

**Text S3, Figure SF2**

Figure SF2 shows the effects of variations in the current data from the four spacecraft on the topology of the reconstruction of Figure 4 in the main text. Panel (a) reduces the current of MMS1(black) and MMS4(blue) by 5%, leaving the others unchanged. There is no discernible difference to Figure 4. Panel (b) reduces MMS1

and MMS4 by 10% with only a slight difference. Panel (c) changes only MMS3(green), increasing the current by 30%, which induces a flux-rope configuration around that spacecraft but the overall topology is still very similar.

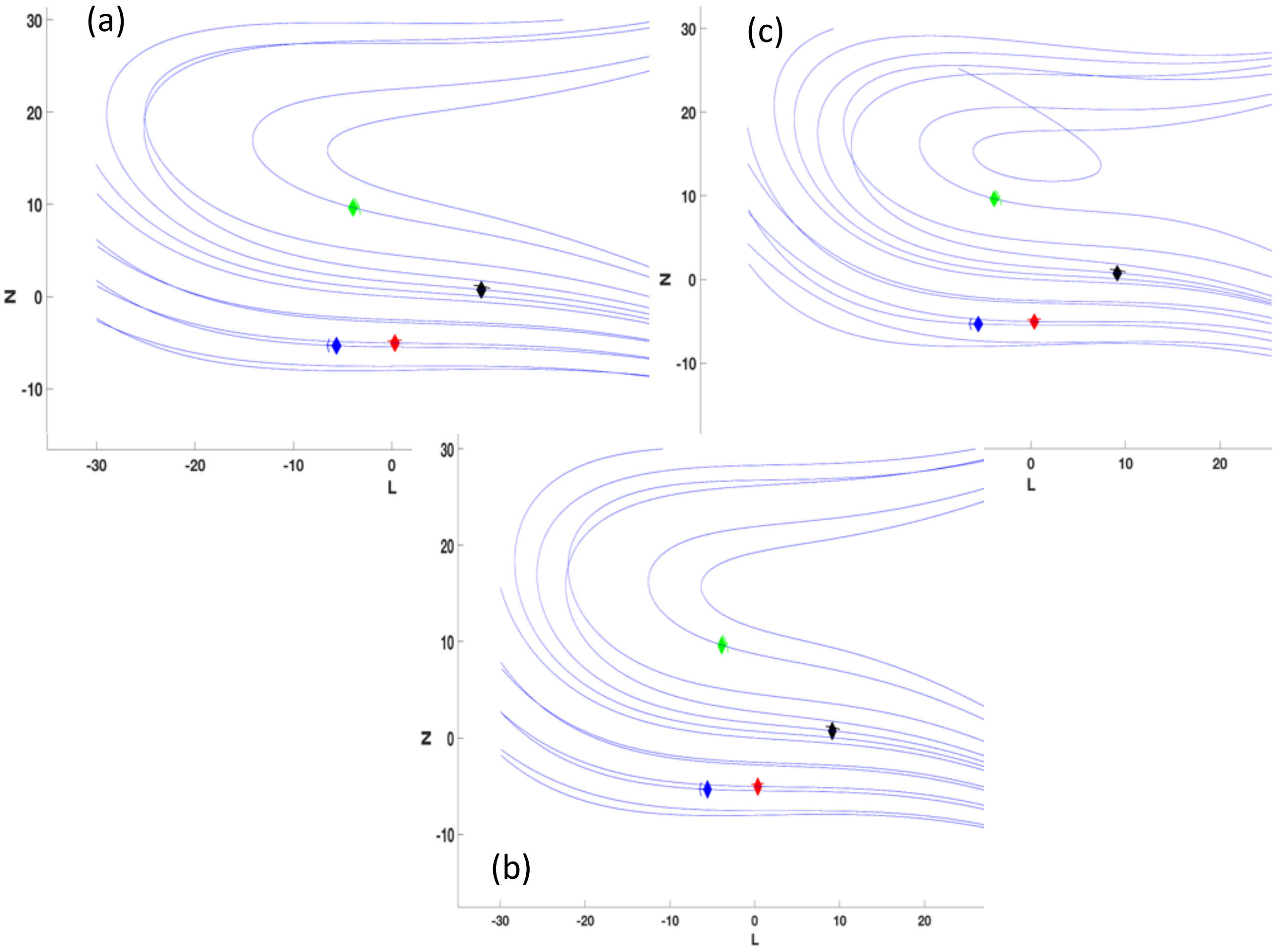

**Figure SF2**

Reconstructions of the data seen in Figure 4, main text, with changes in current values to investigate the effects on topology. In panel (a), the current on the four MMS[1,2,3,4] spacecraft were multiplied by [ 0.95,1.0,1.0,0.95]. In the panel (b), by [ 0.9,1.0,1.0,0.9]. In the panel (c), by [ 1.0,1.0,1.3,1.0].

**V25 version with Div B** **1132** Format1

| | | | | | | | | | | | | | |
|---|---|---|---|---|---|---|---|---|---|---|---|---|---|
| | | | | | | "-T122-T133' | T211 | T311 | 0 | T321 | | | |
| | | | gB11 | gb21 | gb31 | T111 | T121 | T131 | T221 | T231 | T331 | | gB12 |
| | | Bx0 | dxBx | dyBx | dzBx | dxdxBx | dxdyBx | dxdzBx | dydyBx | dydzBx | dzdzBx | By0 | dxBy |
| | | | | | | X | | | X | | | | |
| | | 1 | 2 | 3 | 4 | | 5 | 6 | | 7 | 8 | 9 | 10 |
| 1Bx | 1 | 1 | 1X | 1Y | 1Z | | 1X*1Y | 1X*1Z | | 1Y*1Z | 1Z*1Z/2 | | |
| 1By | 2 | | -1Y | | | | | | | | | 1 | 1X |
| 1Bz | 3 | | | | | | -1Y*1Z | -1Z*1Z/2 | | | | | |
| 1Jx | 4 | | | | | | -1Z | | | | | | |
| 1Jy | 5 | | | | 1 | | | 1X | | 1Y | 1Z | | |
| 1Jz | 6 | | | -1 | | | -1X | | | -1Z | | | 1 |
| 2Bx | 7 | 1 | 2X | 2Y | 2Z | | 2X*2Y | 2X*2Z | | 2Y*2Z | 2Z*2Z/2 | | |
| 2By | 8 | | -2Y | | | | | | | | | 1 | 2X |
| 2Bz | 9 | | | | | | -2Y*2Z | "-2Z*2Z/2" | | | | | |
| 2Jx | 10 | | | | | | -2Z | | | | | | |
| 2Jy | 11 | | | | 1 | | | 2X | | 2Y | 2Z | | |
| 2Jz | 12 | | | -1 | | | -2X | | | -2Z | | | 1 |
| 3Bx | 13 | 1 | 3X | 3Y | 3Z | | 3X*3Y | 3X*3Z | | 3Y*3Z | 3Z*3Z/2 | | |
| 3By | 14 | | -3Y | | | | | | | | | 1 | 3X |
| 3Bz | 15 | | | | | | "-3Y*3Z" | "-3Z*3Z/2" | | | | | |
| 3Jx | 16 | | | | | | "-3Z" | | | | | | |
| 3Jy | 17 | | | | 1 | | | 3X | | 3Y | 3Z | | |
| 3Jz | 18 | | | -1 | | | -3X | | | -3Z | | | 1 |
| 4Bx | 19 | 1 | 4X | 4Y | 4Z | | 4X*4Y | 4X*4Z | | 4Y*4Z | 4Z*4Z/2 | | |
| 4By | 20 | | -4Y | | | | | | | | | 1 | 4X |
| 4Bz | 21 | | | | | | "-4Y*4Z" | "-4Z*4Z/2" | | | | | |
| 4Jx | 22 | | | | | | "-4Z" | | | | | | |
| 4Jy | 23 | | | | 1 | | | 4X | | 4Y | 4Z | | |
| 4Jz | 24 | | | -1 | | | -4X | | | -4Z | | | 1 |
| divB | 25 | | 1 | | | | | | | | | | |

| | | | | | | | | | | | |
|---|---|---|---|---|---|---|---|---|---|---|---|
| | 1 | 2 | 3 | 4 | | 5 | 6 | | 7 | 8 | 9 | 10 |
| gB11 | | 1 | | | | Y | Z | | | | | |
| gB12 | | | | | | | | | | | | 1 |
| gB13 | | | | | | | | | | | | |
| gB21 | | | 1 | | | X | | | Z | | | |
| gB22 | | | | | | | | | | | | |
| gB23 | | | | | | "-Z" | | | | | | |
| gB31 | | | | 1 | | | X | | Y | Z | | |
| gB32 | | | | | | | | | | | | |
| gB33 | | | | | | -Y" | "-Z" | | | | | |
| | | T212 | T312 | 0 | T322 | | | | | | | T213 |
| gb32 | T112 | T122 | T132 | T222 | T232 | T332 | | gB13 | gb23 | gb33 | T113 | T123 |
| dzBy | dxdxBy | dxdyBy | dxdzBy | dydyBy | dydzBy | dzdzBy | Bz0 | dxBz | dyBz | dzBz | dxdxBz | dxdyBz |
| | | | | X | | | | | | | | |
| 11 | 12 | 13 | 14 | | 15 | 16 | 17 | 18 | 19 | 20 | 21 | 22 |
| | | -1X*1X/2 | | | | | | | | | | |
| 1Z | 1X*1X/2 | 1X*1Y | 1X*1Z | | 1Y*1Z | 1Z*1Z/2 | | | | | | |
| | | | | | "-1Z*1Z/2" | | 1 | 1X | 1Y | 1Z | 1X*1X/2 | 1X*1Y |
| -1 | | | -1X | | -1Y | -1Z | | | 1 | | | 1X |
| | | | | | | | | -1 | | | -1X | -1Y |
| | 1X | 1Y | 1Z | | | | | | | | | |
| | | "-2X*2X/2" | | | | | | | | | | |
| 2Z | 2X*2X/2 | 2X*2Y | 2X*2Z | | 2Y*2Z | 2Z*2Z/2 | | | | "-2Y" | | |
| | | | | | "-2Z*2Z/2" | | 1 | 2X | 2Y | 2Z | 2X*2X/2 | 2X*2Y |
| -1 | | | -2X | | -2Y | -2Z | | | 1 | | | 2X |
| | | | | | | | | -1 | | | -2X | -2Y |
| | 2X | 2Y | 2Z | | | | | | | | | |
| | | "-3X*3X/2" | | | | | | | | | | |
| 3Z | 3X*3X/2 | 3X*3Y | 3X*3Z | | 3Y*3Z | 3Z*3Z/2 | | | | "-3Y" | | |
| | | | | | "-3Z*3Z/2" | | 1 | 3X | 3Y | 3Z | 3X*3X/2 | 3X*3Y |
| -1 | | | -3X | | -3Y | -3Z | | | 1 | | | 3X |
| | | | | | | | | -1 | | | -3X | -3Y |
| | 3X | 3Y | 3Z | | | | | | | | | |
| | | "-4X*4X/2" | | | | | | | | | | |
| 4Z | 4X*4X/2 | 4X*4Y | 4X*4Z | | 4Y*4Z | 4Z*4Z/2 | | | | "-4Y" | | |
| | | | | | "-4Z*4Z/2" | | 1 | 4X | 4Y | 4Z | 4X*4X/2 | 4X*4Y |
| -1 | | | -4X | | -4Y | -4Z | | | 1 | | | 4X |
| | | | | | | | | -1 | | | -4X | -4Y |
| | 4X | 4Y | 4Z | | | | | | | | | |
| | | | | | | | | | | 1 | | |
| 11 | 12 | 13 | 14 | | 15 | 16 | 17 | 18 | 19 | 20 | 21 | 22 |
| | | "-X" | | | | | | | | | | |
| | X | Y | Z | | | | | | | | | |
| | | | | | | | | 1 | | | X | Y |
| | | X | | | Z | | | | | | | |
| | | | | | | | | | 1 | | | X |
| 1 | | | X | | Y | Z | | | | | | |
| | | | | | "-Z" | | | | | 1 | | |

| | | | | | | | |
|---|---|---|---|---|---|---|---|
| | | "-T211" | | Format1 | | | |
| T313 | | T323 | "-T311-T322" | | | | |
| T133 | T223 | T233 | T333 | T1132 | gb22 | | |
| dxdzBz | dydyBz | dydzBz | dzdzBz | dxdxdzBy | dyBy | | |
| | X | X | X | | X | | |
| 23 | | | | 24 | 25 | | |
| "-1X*1X/2" | | | | | | 1Bx | 1 |
| | | | | 1X*1X*1Z | 1Y | 1By | 2 |
| 1X*1Z | | | | | | 1Bz | 3 |
| | | | | "-1X*1X" | | 1Jx | 4 |
| -1Z | | | | | | 1Jy | 5 |
| | | | | "2*1X*1Z" | | 1Jz | 6 |
| "-2X*2X/2" | | | | | | 2Bx | 7 |
| | | | | 2X*2X*2Z | 2Y | 2By | 8 |
| 2X*2Z | | | | | | 2Bz | 9 |
| | | | | "-2X*2X" | | 2Jx | 10 |
| -2Z | | | | | | 2Jy | 11 |
| | | | | "2*2X*2Z" | | 2Jz | 12 |
| "-3X*3X/2" | | | | | | 3Bx | 13 |
| | | | | 3X*3X*3Z | 3Y | 3By | 14 |
| 3X*3Z | | | | | | 3Bz | 15 |
| | | | | "-3X*3X" | | 3Jx | 16 |
| -3Z | | | | | | 3Jy | 17 |
| | | | | "2*3X*3Z" | | 3Jz | 18 |
| "-4X*4X/2" | | | | | | 4Bx | 19 |
| | | | | 4X*4X*4Z | 4Y | 4By | 20 |
| 4X*4Z | | | | | | 4Bz | 21 |
| | | | | "-4X*4X" | | 4Jx | 22 |
| -4Z | | | | | | 4Jy | 23 |
| | | | | "2*4X*4Z" | | 4Jz | 24 |
| | | | | | 1 | divB | 25 |

| | | | |
|---|---|---|---|
| 23 | 24 | 25 | |
| "-X" | | | gB11 |
| | 2*X*Z | | gB12 |
| Z | | | gB13 |
| | | | gB21 |
| | | 1 | gB22 |
| | | | gB23 |
| | | | gB31 |
| | X*X | | gB32 |
| X | | | gB33 |